# Higher-order interactions in quantum optomechanics: Revisiting theoretical foundations

**Sina Khorasani** [1,2]

[1] School of Electrical Engineering, Sharif University of Technology, Tehran, Iran; khorasani@sina.sharif.edu
[2] École Polytechnique Fédérale de Lausanne, Lausanne, CH-1015, Switzerland; sina.khorasani@epfl.ch

**Abstract:** The theory of quantum optomechanics is reconstructed from first principles by finding a Lagrangian from light's equation of motion and then proceeding to the Hamiltonian. The nonlinear terms, including the quadratic and higher-order interactions, do not vanish under any possible choice of canonical parameters, and lead to coupling of momentum and field. The existence of quadratic mechanical parametric interaction is then demonstrated rigorously, which has been so far assumed phenomenologically in previous studies. Corrections to the quadratic terms are particularly significant when the mechanical frequency is of the same order or larger than the electromagnetic frequency. Further discussions on the squeezing as well as relativistic corrections are presented.

**Keywords:** Optomechanics, Quantum Physics, Nonlinear Interactions

## 1. Introduction

The general field of quantum optomechanics is based on the standard optomechanical Hamiltonian, which is expressed as the simple product of photon number $\hat{n}$ and the position $\hat{x}$ operators, having the form $\mathbb{H}_\text{OM} = \hbar g_0 \hat{n}\hat{x}$ [1-4] with $g_0$ being the single-photon coupling rate. This is mostly referred to a classical paper by Law [5], where the non-relativistic Hamiltonian is obtained through Lagrangian dynamics of the system. This basic interaction is behind numerous exciting theoretical and experimental studies, which demonstrate a wide range of applications. The optomechanical interaction $\mathbb{H}_\text{OM}$ is inherently nonlinear by its nature, which is quite analogous to the third-order Kerr optical effect in nonlinear optics [6,7]. These for instance include the optomechanical arrays [8-17], squeezing of phonon states [18-20], Heisenberg's limited measurements [21], non-reciprocal optomechanical systems [22-28], sensing [29-31], engineered dissipation [32], engineered states [33], and non-reciprocal acousto-optical effects in optomechanical crystals [34-36].

Recent ideas in this field such as microwave-optical conversion [37-40], cavity electrooptics [41-43], optomechanical induced transparency [44-47], and optomechanical verification of Bell's inequality [48] all emerge as closely related duals of quantum optomechanical systems, which are described within a completely identical framework. Furthermore, quantum chaos which had been predicted in quantum optomechanics [49,50] and cavity quantum electro-dynamical systems [51,52], has been recently observed in optomechanics [53,54].

Usually, the analysis in all these above works is ultimately done within the linearized approximation of ladder operators, thus being limited in accuracy where $\mathbb{H}_\text{OM}$ interaction is non-existent or vanishingly small. For applications where quadratic or even quartic effects are primarily pursued, $g_0$ may be designed to be identically zero [55-61], which urges need for accurate knowledge of higher-order interaction terms. Similar situation also could arise in trapped ultracold atomic gases [62], where linear interactions identically vanish. Relevant Physical phenomena in optomechanics such as four-wave mixing, also is suitably described by higher order interaction terms [63]. Moreover, significance and prominent role of such nonlinear interactions has been observed in few recent experiments [64,65].

A careful review of the theory of this subject [5], however, reveals that there are a number of physical approximations in formulation of the problem such as the non-relativistic limit [66,67], which makes the unified description of relativistic photon momenta and non-relativistic mirror motion inaccurate. As it is being shown here, a full treatment of the latter will yield higher-order multi-particle interactions. Such quadratic interactions have been recently used phenomenologically [58] without a theoretical basis. Similar yet weaker interactions may be also drawn from relativistic corrections [68] as discussed here. In that sense, the quadratic interactions are shown to receive contributions from both non-relativistic and relativistic terms, which become quite significant when the mechanical frequency is comparable or larger than the electromagnetic frequency.

## 2. Classical Hamiltonian

The basic theory to be discussed is based on two important and basic assumptions. We first assume that the cavity mode decomposition is valid independent of the mirror motion. Second, the electric fields vanish at the mirror surface in the frame that the Lagrangian is constructed. These assumptions seem reasonable in the usual discussions with much lower mechanical frequencies when the cavity mode change can be treated adiabatically. However, when the end mirror oscillates at a very high frequency, it dresses the cavity modes so that the mode frequency might become undefined invalidating the assumption of mode decomposition. At higher frequencies, it has also been known and demonstrated that the mirror undergoing relativistic motion could produce photons from vacuum, known as the dynamical Casimir effect [69,70]. This is of course beyond the regime being explored in this article.

The focus of the first two subsections 2.1 and 2.2 is to assert the claim, and demonstrate what term is missing and why it happens. As it will be shown and rigorously proven, even for the simplest case of interaction with a single-optical mode, a new term of the type $\dot{q}^2 Q^2$ representing quadratic momentum $\dot{q}$ and optical field $Q$ interactions is found, the origin of which is also identified. For the more general case of multi-mode optical fields, the situation is even much more complex and there are a few more missing terms to consider. Once the Lagrangian is known, the Hamiltonian is subsequently constructed in the subsection 2.3.

### 2.1. The Equations of Motion

The one-dimensional wave equation for transverse component of the magnetic potential $A(x,t)$ in the dimensional form is expressed as [5]

$$c^2 A_{xx}(x,t) = A_{tt}(x,t), \tag{1}$$

where $x$ and $t$ are respectively the position and time coordinates, and $c$ is the speed of light in free space. Suppose the Fourier series relations for the magnetic vector potential are defined as [5]

$$Q_k(t) = \frac{1}{c}\sqrt{\frac{2S}{\mu_0 q(t)}} \int_0^{q(t)} A(x,t)\sin[\kappa_k(t)x]dx,$$

$$A(x,t) = c\sqrt{\frac{2\mu_0}{Sq(t)}} \sum_{k=1}^{\infty} Q_k(t)\sin[\kappa_k(t)x],$$

$$\omega_k(t) = c\pi k/q(t) = c\kappa_k(t), \tag{2}$$

where $S$ is the cross-sectional area, $\mu_0$ is the permeability of vacuum. This arrangement ensures that the definition of canonical variables can be used later, so that $\dot{Q}_k^2$ simply takes on the dimension of energy.

One may furthermore define the functions $f_k = \sqrt{2/q}\sin(\kappa_k x)$ and $g_k = \sqrt{2/q}\cos(\kappa_k x)$, and hence $A(x,t) = s\sum Q_k f_k$ where $s = c\sqrt{\mu_0/S}$. Here, the inner product is also defined as $(a|b) = \int_0^q ab\,dx$ such that the following relations may be found

$$\left(f_k\middle|f_j\right) = \delta_{kj}, \qquad \left(f_k\middle|\kappa_j x\middle|g_j\right) = \alpha_{kj}, \qquad \left(f_k\middle|\kappa_j^2 x^2\middle|g_j\right) = \beta_{kj}. \tag{3}$$

After straightforward calculations one obtains

$$\alpha_{kj} = -\frac{1}{2}\delta_{kj} + g_{kj},$$

$$\beta_{kj} = \left(k^2\frac{\pi^2}{3} - \frac{1}{2}\right)\delta_{kj} + h_{kj} = \begin{cases} k^2\dfrac{\pi^2}{3} - \dfrac{1}{2}, & k = j, \\ 8\dfrac{(-1)^{k+j}kj^3}{(k^2 - j^2)^2}, & k \neq j. \end{cases} \tag{4}$$

Here, the anti-symmetric coefficients $g_{kj}$ and $h_{kj}$ are

$$g_{kj} = -g_{jk} = \begin{cases} 0, & k = j, \\ 2\dfrac{(-1)^{k+j}kj}{j^2 - k^2}, & k \neq j, \end{cases} \tag{5}$$

$$h_{kj} = \begin{cases} 0, & k = j, \\ 8\dfrac{(-1)^{k+j}kj^3}{(k^2 - j^2)^2}, & k \neq j. \end{cases}$$

Differentiating $f_k$ and $g_k$ with respect to $t$, noting that $\dot{\kappa}_k = -\dot{q}\kappa_k/q$, gives

$$\dot{f}_k = -\frac{\dot{q}}{q}\left(\frac{1}{2}f_k + x\kappa_k g_k\right),$$

$$\dot{g}_k = -\frac{\dot{q}}{q}\left(\frac{1}{2}g_k - x\kappa_k f_k\right),$$

$$\ddot{f}_k = -\frac{\ddot{q}q - \dot{q}^2}{q^2}\left(\frac{1}{2}f_k + x\kappa_k g_k\right) + \frac{\dot{q}^2}{q^2}\left(\frac{1}{4}f_k + 2x\kappa_k g_k - x^2\kappa_k^2 f_k\right). \tag{6}$$

It is possible now to differentiate $A(x,t)$ with respect to position and time to obtain the relations

$$\frac{c^2}{s}A_{xx}(x,t) = -\sum \omega_k^2 Q_k f_k,$$

$$\frac{1}{s}A_t(x,t) = \sum \dot{Q}_k f_k - \frac{\dot{q}}{q}\sum Q_k\left(\frac{1}{2}f_k + x\kappa_k g_k\right),$$

$$\frac{1}{s}A_{tt}(x,t) = \sum \ddot{Q}_k f_k - 2\frac{\dot{q}}{q}\sum \dot{Q}_k\left(\frac{1}{2}f_k + x\kappa_k g_k\right) - \frac{\ddot{q}q - \dot{q}^2}{q^2}\sum Q_k\left(\frac{1}{2}f_k + x\kappa_k g_k\right)$$
$$+ \frac{\dot{q}^2}{q^2}\sum Q_k\left(\frac{1}{4}f_k + 2x\kappa_k g_k - x^2\kappa_k^2 f_k\right). \tag{7}$$

Therefore, the wave equation reads

$$-\sum \omega_k^2 Q_k f_k = \sum \ddot{Q}_k f_k - \frac{\dot{q}}{q}\sum \dot{Q}_k(f_k + 2x\kappa_k g_k) - \frac{\ddot{q}}{q}\sum Q_k\left(\frac{1}{2}f_k + x\kappa_k g_k\right)$$
$$+ \frac{\dot{q}^2}{q^2}\sum Q_k\left(\frac{3}{4}f_k + 3x\kappa_k g_k - x^2\kappa_k^2 f_k\right). \tag{8}$$

Now, the inner product relationships (3) and some further simplification (Appendix A) help us to obtain the equation of motion as

$$\ddot{Q}_k = -\omega_k^2 Q_k + r_k\frac{\dot{q}^2}{q^2}Q_k + 2\frac{\dot{q}}{q}\sum g_{kj}\dot{Q}_j + \frac{\ddot{q}}{q}\sum g_{kj}Q_j + \frac{\dot{q}^2}{q^2}\sum\left(h_{kj} - 3g_{kj}\right)Q_j, \tag{9}$$

where all summations are nonzero only for $j \neq k$, and

$$r_k = k^2\frac{\pi^2}{3} + \frac{1}{4}. \tag{10}$$

The related equation in Law's paper [5]

$$\ddot{Q}_k = -\omega_k^2 Q_k + 2\frac{\dot{q}}{q}\sum g_{kj}\dot{Q}_j + \frac{\ddot{q}}{q}\sum g_{kj}Q_j + \frac{\dot{q}^2}{q^2}\sum g_{kj}g_{lj}Q_l - \frac{\dot{q}^2}{q^2}\sum g_{kj}Q_j, \tag{11}$$

is completely equivalent, but apparently different in presentation. Similarly, one may directly deduce from the Newton's equation of motion

$$m\ddot{q} = -\frac{\partial}{\partial x}V(x) + \frac{S}{2\mu_0}B^2(q,t) = -\frac{\partial}{\partial x}V(x) + \frac{1}{q}\sum(-1)^{k+j}\omega_k\omega_j Q_k Q_j. \tag{12}$$

*2.2. Lagrangian*

The associated Lagrangian which leads to the above set of Euler equations is given by

$$\mathcal{L} = \frac{1}{2}m\dot{q}^2 - V(q) + \frac{1}{2}\sum\left(\dot{Q}_k^2 - \omega_k^2 Q_k^2\right) - \frac{\dot{q}}{q}\sum g_{kj}Q_j\dot{Q}_k$$
$$+ \frac{\dot{q}^2}{2q^2}\sum(h_{kj} - 2g_{kj} + r_k\delta_{kj})Q_k Q_j. \tag{13}$$

It should be again noticed that $h_{kk} = 0$ by (5), which together with the antisymmetry of coefficients $g_{kj}$ from (5), allows further simplification to reach

$$\mathcal{L} = \frac{1}{2}m\dot{q}^2 - V(q) + \frac{1}{2}\sum\left(\dot{Q}_k^2 - \omega_k^2 Q_k^2\right) + \frac{\dot{q}^2}{2q^2}\sum d_{kj}Q_k Q_j - \frac{\dot{q}}{q}\sum g_{kj}Q_j \dot{Q}_k, \tag{14}$$

where $d_{kj} = \frac{1}{2}(h_{kj} + h_{jk}) + r_k \delta_{kj}$ is related to the symmetric part of $h_{kj}$, given by

$$d_{kj} = d_{jk} = \begin{cases} r_k, & k = j, \\ 4\dfrac{(-1)^{k+j}kj(k^2+j^2)}{(k^2-j^2)^2}, & k \neq j. \end{cases} \tag{15}$$

This is to be compared with the quite different expression by Law [5] given by

$$\mathcal{L} = \frac{1}{2}m\dot{q}^2 - V(q) + \frac{1}{2}\sum\left(\dot{Q}_k^2 - \omega_k^2 Q_k^2\right) + \frac{\dot{q}^2}{2q^2}\sum g_{lk}g_{lj}Q_k Q_j - \frac{\dot{q}}{q}\sum g_{kj}Q_j \dot{Q}_k, \tag{16}$$

which indeed consistently satisfies the Euler's pair of equations.

Now, it is a matter of speculation whether (9) is equivalent to (11) or not. Firstly, the diagonal terms are equal if

$$\sum_k g_{kj}g_{kj} = r_j = d_{jj}, \tag{17}$$

which holds true because of the identity

$$\sum_{j=1, j \neq k}^{\infty} \frac{4k^2 j^2}{(j^2 - k^2)^2} = k^2 \frac{\pi^2}{3} + \frac{1}{4}. \tag{18}$$

Now, for a single-mode system with only one radiation mode, we may easily notice that (9) and (11) become readily identical. This becomes more clear by noticing that $g_{kk} = h_{kk} = 0$, and hence for a system with only one electromagnetic radiation mode, (9) becomes $\ddot{Q} = -\omega^2 Q + (rq^{-2}\dot{q}^2)Q$, with $r = r_1$. This is while Law's expression (11) is $\ddot{Q} = -\omega^2 Q + q^{-2}\dot{q}^2\left(\sum g_{1j}g_{1j}\right)Q$. But (17) requires that $r = \sum g_{1j}g_{1j}$, which confirms the equivalency for single-mode systems.

As for a multimode system, and by comparing (13) and (16) one would need

$$\sum g_{kl}g_{jl} = d_{kj} = r_k \delta_{kj} - 2g_{kj} + h_{kj}, \tag{19}$$

in order to (9) and (11) be identical. This can be put to numerical tests (here done by the author by coding a simple Mathematica program), and is in fact accurately satisfied. Hence, the single-mode Lagrangian in the non-relativistic limit can be written in the form

$$\mathcal{L} = \frac{1}{2}m\dot{q}^2 - V(q) + \frac{1}{2}[\dot{Q}^2 - \omega^2(q)Q^2] + \frac{\dot{q}^2}{2q^2}rQ^2. \tag{20}$$

The last term has been usually ignored so far in the literature, and will result in momentum-field coupling. In the remainder of the paper, we focus on nonlinear terms arising from this interaction, and then also add up the relativistic corrections in the end.

*2.3. Hamiltonian*

The definition of canonical momenta taken here is

$$P_k = \dot{Q}_k - \frac{\dot{q}}{q}\sum A_{kj}Q_j, \qquad p = m\dot{q} - \frac{1}{q}\sum B_{kj}P_k Q_j, \tag{21}$$

with $A_{kj}$ and $B_{kj}$ being some transformation coefficients to be determined later. One here may take advantage of the degree of freedom in choice of $A_{kj}$ and $B_{kj}$ to get rid of unwanted summation terms in the Hamiltonian. It has to be here noticed again that the existence of the last term of (20), being completely new even under the single-mode operation, has nothing to do with the choice of canonical momenta. That implies the final resulting single-mode (and therefore multi-mode) Hamiltonian will be inevitably different, incorporating a few new terms. The Hamiltonian may be now derived from the Lagrangian by iterated use of (21) and through the relationship $\mathcal{H} = p\dot{q} + \sum P_k \dot{Q}_k - \mathcal{L}$ as

$$\mathcal{H} = \left[m\dot{q} - \frac{1}{q}\sum B_{kj}\left(\dot{Q}_k - \frac{\dot{q}}{q}\sum A_{kl}Q_l\right)Q_j\right]\dot{q} + \sum\left(\dot{Q}_k - \frac{\dot{q}}{q}\sum A_{kj}Q_j\right)\dot{Q}_k - \mathcal{L}. \tag{22}$$

Law [5] arbitrates the choice $A_{kj} = B_{kj} = g_{kj}$. But by going further with this choice for a single-mode optical field, it will be evident that $P = \dot{Q}$ and $p = m\dot{q}$, hence, still resulting in an extra nonlinear term proportional to $p^2Q^2/q^2$ in the Hamiltonian, leading to a fourth-order momentum-field interaction. This will be discussed shortly in the subsection below.

Hence, the Lagrangian $\mathcal{L}$ found in the above yields the Hamiltonian below after some algebra

$$\mathcal{H} = \frac{1}{2}m\dot{q}^2 + V(q) + \frac{1}{2}\sum\left(\dot{Q}_k^2 + \omega_k^2 Q_k^2\right) - \frac{\dot{q}}{q}\sum(A_{kj} + B_{kj} - g_{kj})\dot{Q}_k Q_j$$
$$+ \frac{\dot{q}^2}{2q^2}\sum\left(2\sum B_{kj}A_{kl} - d_{jl}\right)Q_l Q_j. \tag{23}$$

Here is readily evident now by (19), that the last term can be made identically zero, only if $d_{jl} = \sum g_{kj}g_{kl} = 2\sum B_{kj}A_{kl}$. This not only cannot be satisfied by Law's choice $A_{kj} = B_{kj} = g_{kj}$, but also the second summation term nonlinear in $\dot{q}\dot{Q}_k$ will also survive, further complicating the Hamiltonian formulation.

Now, further elimination of $\dot{q}$ and $\dot{Q}_k$ from (21) gives

$$\mathcal{H} = \frac{1}{2m}\left(p + \frac{1}{q}\sum B_{kj}P_k Q_j\right)^2 + V(q) + \frac{1}{2}\sum \omega_k^2 Q_k^2$$
$$+ \frac{1}{2}\sum\left[P_k + \frac{1}{mq}\left(p + \frac{1}{q}\sum B_{lm}P_l Q_m\right)\sum A_{kj}Q_j\right]^2$$
$$- \frac{1}{mq}\left(p + \frac{1}{q}\sum B_{kj}P_k Q_j\right)\sum C_{kj}\left[P_k + \frac{1}{mq}\left(p + \frac{1}{q}\sum B_{kj}P_k Q_j\right)\sum A_{kj}Q_j\right]Q_j$$
$$+ \frac{1}{2m^2q^2}\left(p + \frac{1}{q}\sum B_{kj}P_k Q_j\right)^2 \sum D_{jl}Q_l Q_j. \tag{24}$$

Here, $C_{kj} = A_{kj} + B_{kj} - g_{kj}$ and $D_{jl} = 2\sum B_{kj}A_{kl} - d_{jl}$. Dealing directly with such an intractable and long expression is without doubt too tough. Instead, we may tweak Law's choice slightly as $A_{kj} = B_{kj} = \frac{1}{2}g_{kj}$, which allows (23) be greatly simplified as

$$\mathcal{H} = \frac{1}{2}m\dot{q}^2 + V(q) + \frac{1}{2}\sum\left(\dot{Q}_k^2 + \omega_k^2 Q_k^2\right) - \frac{\dot{q}^2}{4q^2}\sum d_{kj}Q_k Q_j. \tag{25}$$

This can be further eventually expanded and simplified as

$$\mathcal{H} = \frac{1}{2m}\left(p + \frac{1}{2q}\sum g_{kj}P_k Q_j\right)^2 + V(q) + \frac{1}{2}\sum\left(P_k^2 + \omega_k^2 Q_k^2\right)$$
$$+ \frac{1}{4mq}\left(p + \frac{1}{2q}\sum g_{kj}P_k Q_j\right)\sum g_{kj}P_k Q_j - \frac{1}{8m^2q^2}\left(p + \frac{1}{2q}\sum g_{kj}P_k Q_j\right)^2 \sum d_{kj}Q_k Q_j. \tag{26}$$

The last two terms of this Hamiltonian can be expanded to obtain multiple orders of interactions. These include higher-order tripartite phonon/two-photon, and quadpartite two-phonon/two-photon interactions, which does not exist in the Law's Hamiltonian [5], given by

$$\mathcal{H} = \frac{1}{2m}\left(p + \frac{1}{q}\sum g_{kj}P_k Q_j\right)^2 + V(q) + \frac{1}{2}\sum\left(P_k^2 + \omega_k^2 Q_k^2\right). \tag{27}$$

As it appears, (27) is missing two very different types of momentum field interaction as the last two summation terms of (26). This fact becomes evident in below.

### 2.3.1. Single Optical Mode

The interesting difference between these two Hamiltonians becomes quite clear with consideration of only one optical mode in the cavity. This simplifies our derived Hamiltonian to

$$\mathcal{H} = \frac{1}{2m}p^2 + V(q) + \frac{1}{2}\left[P^2 + \frac{\pi^2}{q^2}Q^2\right] - \frac{r}{8m^2q^2}p^2Q^2, \tag{28}$$

where $\omega(q) = \pi/q$ and $r \cong 3.8$, while the Law's Hamiltonian [5] gives rise to the significantly different form
5

$$\mathcal{H} = \frac{1}{2m}p^2 + V(q) + \frac{1}{2}\left[P^2 + \frac{\pi^2}{q^2}Q^2\right]. \tag{29}$$

As it will be shown below, (28) and (29) agree only to the first order, and hence up to the standard optomechanical Hamiltonian.

### 3. Field Quantization

When the obtained Hamiltonian is moved to the realm of quantum mechanics, it is first needed to define the non-commutation rules $[\hat{q},\hat{p}] = i\hbar$ and $[\hat{Q}_k,\hat{P}_j] = i\hbar\delta_{kj}$, with the commutation rules $[\hat{q},\hat{Q}_k] = [\hat{Q}_k,\hat{p}] = [\hat{q},\hat{P}_j] = [\hat{P}_j,\hat{p}] = 0$. This allows us to introduce the field creation and annihilation operators according to

$$\hat{Q}_k = \sqrt{\frac{\hbar}{2\omega_k(\hat{q})}}\left(\hat{a}_k^\dagger + \hat{a}_k\right) = \sqrt{\frac{\hbar}{\omega_k(\hat{q})}}\mathbb{Q}_k,$$

$$\hat{P}_k = i\sqrt{\frac{\hbar\omega_k(\hat{q})}{2}}\left(\hat{a}_k^\dagger - \hat{a}_k\right) = \sqrt{\hbar\omega_k(\hat{q})}\mathbb{P}_k, \tag{30}$$

where $\omega_k(\hat{q}) = c\pi k/\hat{q}$ is defined according to (2). Also for a mechanical resonant frequency $\Omega$ and a spring restoring potential

$$V(\hat{q}) = \frac{1}{2}m\Omega^2(\hat{q}-l)^2, \tag{31}$$

the displacement operator may be defined as $\hat{q} = l + \hat{x}$, where $l$ is the reference position of mirror, and hence the phonon ladder operators as

$$\hat{x} = \sqrt{\frac{\hbar}{2m\Omega}}\left(\hat{b}^\dagger + \hat{b}\right) = \sqrt{\frac{\hbar}{m\Omega}}\mathfrak{x}, \qquad \hat{p} = i\sqrt{\frac{\hbar m\Omega}{2}}\left(\hat{b}^\dagger - \hat{b}\right) = \sqrt{\hbar m\Omega}\mathfrak{P}, \tag{32}$$

with $[\hat{b},\hat{b}^\dagger] = 1$ and $[\mathfrak{x},\mathfrak{P}] = i$.

Now, it is necessary first to symmetrize [67] the classical Hamiltonian prior to insertion of operators, to ensure correct quantization of parameters. The process of symmetrization is done according to [71-73]

$$\mathcal{S}\{\mathbb{AB}\} = \frac{1}{2}(\mathbb{AB}+\mathbb{BA}), \qquad \mathcal{S}\{\mathbb{ABC}\} = \frac{1}{3}(\mathbb{A}\mathcal{S}\{\mathbb{BC}\}+\mathbb{B}\mathcal{S}\{\mathbb{AC}\}+\mathbb{C}\mathcal{S}\{\mathbb{AB}\}), \tag{33}$$

etc. Therefore, after symmetrization the final form of the Hamiltonian is given by

$$\mathbb{H} = \frac{1}{2m}\hat{p}^2 + V(\hat{q}) + \frac{1}{2}\sum\left(\hat{P}_k^2 + \omega_k^2\hat{Q}_k^2\right) + \frac{1}{4m}\mathcal{S}\left\{\frac{1}{\hat{q}^2}\left(\hat{p}\hat{q}+\frac{1}{2}\sum g_{kj}\hat{P}_k\hat{Q}_j\right)\sum g_{kj}\hat{P}_k\hat{Q}_j\right\}$$
$$-\frac{r}{8m^2}\mathcal{S}\left\{\frac{1}{\hat{q}^4}\left(\hat{p}\hat{q}+\frac{1}{2}\sum g_{kj}\hat{P}_k\hat{Q}_j\right)^2\sum d_{kj}\hat{Q}_k\hat{Q}_j\right\}. \tag{34}$$

For a single-optical mode, (34) greatly simplifies and one gets

$$\mathbb{H} = \frac{1}{2m}\hat{p}^2 + V(\hat{q}) + \frac{1}{2}\left(\hat{P}^2 + \frac{\pi^2}{\hat{q}^2}\hat{Q}^2\right) - \frac{r}{8m^2}\mathcal{S}\left\{\frac{1}{\hat{q}^2}\hat{p}^2\right\}\hat{Q}^2. \tag{35}$$

This has to be applied to the last interacting term, which involves

$$\mathcal{S}\left\{\frac{1}{\hat{q}^2}\hat{p}^2\right\} = \mathcal{S}\left\{\frac{\hat{p}^2}{\hat{q}^2}\right\}. \tag{36}$$

But symmetrization of a term which contains $n$ non-commuting terms, results in $n!$ terms, which for this case sum up to a total of $4! = 24$ different expressions. The direct way to get around this situation is to first make an estimate of which terms are the strongest in the limit of linearized interaction and ignore the rest. It is possible furthermore to use the approximate replacement

$$\frac{1}{\hat{q}^n} \cong \frac{1}{l^n}\left(1 - n\frac{\hat{x}}{l}\right), \tag{37}$$

to obtain

$$\mathbb{H} = \frac{1}{2m}\hat{p}^2 + V(\hat{q}) + \frac{1}{2}\left[\hat{P}^2 + \omega^2\left(1 - 2\frac{\hat{x}}{l} + \frac{4}{l^2}\hat{x}^2 + \cdots\right)\hat{Q}^2\right] - \frac{r}{8m^2l^2}\mathcal{S}\left\{\hat{p}^2\left(1 - 2\frac{\hat{x}}{l}\right)\right\}\hat{Q}^2, \quad (38)$$

where further substitutions should be taken as

$$\hat{P} \cong \sqrt{\hbar\omega}\left(1 - \frac{1}{2l}\hat{x} + \frac{3}{8l^2}\hat{x}^2 + \cdots\right)\mathbb{P}, \qquad \hat{Q} \cong \sqrt{\frac{\hbar}{\omega}}\left(1 + \frac{1}{2l}\hat{x} - \frac{1}{8l^2}\hat{x}^2 + \cdots\right)\mathbb{Q}. \quad (39)$$

This can be decomposed to the terms

$$\mathbb{H} = \mathbb{H}_{012} + \mathbb{H}_3 + \mathbb{H}_4 + \mathbb{H}_5 + \cdots, \qquad \mathbb{H}_{012} = \frac{1}{2}\hbar\Omega\mathfrak{P}^2 + U(\mathfrak{X}) + \frac{1}{2}\hbar\omega(\mathbb{P}^2 + \mathbb{Q}^2). \quad (40)$$

Hence, there are several distinct types of nonlinear optomechanical multi-phonon/multi-photon interactions, which by defining $R = r/4 \approx 0.95$ are respectively given by

$$\mathbb{H}_3 = -\frac{\hbar\omega}{2l}\sqrt{\frac{\hbar}{m\Omega}}\mathfrak{X}(\mathbb{P}^2 + \mathbb{Q}^2), \qquad \mathbb{H}_4 = \frac{\hbar^2}{2l^2 m}\left[-R\frac{\Omega}{\omega}\mathfrak{P}^2\mathbb{Q}^2 + \frac{\omega}{\Omega}\mathfrak{X}^2(\mathbb{P}^2 + \mathbb{Q}^2)\right],$$

$$\mathbb{H}_5 = -\frac{\hbar^{\frac{5}{2}}}{2m^{\frac{3}{2}}l^3\sqrt{\Omega}}\left\{-2R\frac{\Omega}{\omega}\mathcal{S}\{\mathfrak{P}^2\mathfrak{X}\}\mathbb{Q}^2 + \frac{\omega}{\Omega}\mathfrak{X}^3(\mathbb{P}^2 + \mathbb{Q}^2)\right\}, \quad (41)$$

and so on for higher order interactions. Here, the expansion of symmetrized terms, for instance, gives

$$\mathcal{S}\{\mathfrak{P}^2\mathfrak{X}\} = \frac{1}{3}(\mathfrak{P}^2\mathfrak{X} + \mathfrak{X}\mathfrak{P}^2 + \mathfrak{P}\mathfrak{X}\mathfrak{P}). \quad (42)$$

Now, it is noted that since usually $\omega \gg \Omega$, it may observed that the first terms are much weaker than the second terms. Hence, using the identity $\mathbb{P}^2 + \mathbb{Q}^2 = \frac{1}{2}\hat{n} + \frac{1}{4}$ where $\hat{n}$ is photon number operator, the following is obtained

$$\mathbb{H}_3 = -\frac{\hbar\omega}{l}\sqrt{\frac{\hbar}{m\Omega}}\mathfrak{X}\left(\hat{n} + \frac{1}{2}\right) = -\hbar\alpha\mathfrak{X}\left(\hat{n} + \frac{1}{2}\right),$$

$$\mathbb{H}_4 = \frac{\hbar^2}{l^2 m}\frac{\omega}{\Omega}\mathfrak{X}^2\left(\hat{n} + \frac{1}{2}\right) = +\hbar\beta\mathfrak{X}^2\left(\hat{n} + \frac{1}{2}\right),$$

$$\mathbb{H}_5 = -\frac{\omega}{\Omega}\frac{\hbar^{\frac{5}{2}}}{m^{\frac{3}{2}}l^3\sqrt{\Omega}}\mathfrak{X}^3\left(\hat{n} + \frac{1}{2}\right) = -\hbar\gamma\mathfrak{X}^3\left(\hat{n} + \frac{1}{2}\right), \quad (43)$$

where $\mathbb{H}_3 \equiv \mathbb{H}_{OM}$ is the simple optomechanical interaction, and $\mathbb{H}_4$ is known as the quadratic interaction. It has to be emphasized that while $\mathbb{H}_3 \equiv \mathbb{H}_{OM}$ is actually nonlinear in the exact mathematical sense, it is the quadratic interaction $\mathbb{H}_4$ which is mostly referred to as the nonlinear interaction in the literature [55-57]. Since it is possible to make $g_0$ and therefore $\mathbb{H}_3$ identically vanish by appropriate optomechanical design [55-58] in which the overlap integral of optical and mechanical modes sums up to zero, hence the quadratic interactions $\mathbb{H}_4$ can then find physical significance.

The quadratic interaction has been a subject of growing importance in the recent years in optomechanical systems [59-61] and beyond [62]. In [59] the photon statistics and blockade under $\mathbb{H}_4$ interactions has been studied and analytical expressions were derived. The quantum dissipative master function has been numerically solved and the corresponding correlation functions were obtained. Interestingly, quadratic optomechanical interactions can arise at the single-photon level, too, where rigorous analytical solutions have been devised [60]. Such type of interactions can be also well described using equivalent nonlinear electrical circuits, where a Josephson junction brings in the desired nonlinearity of quadratic interactions and terminate a pair of lumped transmission lines [61]. Finally, ultracold atoms also can exhibit interactions of a comparable type which is mathematically equivalent to the quadratic interaction [62].

The single-photon multi-particle rates are given by

$$\alpha = \frac{\omega}{l}\sqrt{\frac{\hbar}{m\Omega}} \equiv \sqrt{2}g_0, \qquad \beta = \frac{\hbar}{l^2 m}\frac{\omega}{\Omega}, \qquad \gamma = \frac{\omega}{\Omega}\frac{\hbar^{\frac{3}{2}}}{m^{\frac{3}{2}}l^3\sqrt{\Omega}}. \tag{44}$$

This summarizes the Hamiltonian as

$$\mathbb{H} = \mathbb{H}_0 + \mathbb{H}_{int}. \tag{45}$$

in which $\mathbb{H}_0$ and $\mathbb{H}_{int}$ are respectively the non-interacting and interacting Hamiltonians

$$\mathbb{H}_0 = \mathbb{H}_{012} - \frac{1}{2}\hbar(\alpha\mathfrak{X} - \beta\mathfrak{X}^2 + \gamma\mathfrak{X}^3 + \cdots), \qquad \mathbb{H}_{int} = \hbar(\alpha\mathfrak{X} - \beta\mathfrak{X}^2 + \gamma\mathfrak{X}^3 + \cdots)\hat{n}, \tag{46}$$

when $\omega \gg \Omega$. Now, the dimensionless constant $\theta$ is defined as

$$\theta = \frac{1}{l}\sqrt{\frac{\hbar}{m\Omega}} = \frac{x_{zp}}{l}, \tag{47}$$

with $x_{zp}$ being the r.m.s. value of zero-point fluctuations, by which the following is deduced

$$\beta = \theta\alpha \ll \alpha, \qquad \gamma = \theta\beta = \theta^2\alpha \ll \beta. \tag{48}$$

This implies that every kind of higher-order interaction is typically $\theta$ times weaker than the interaction of the preceding-order. It should be noted that while such interactions are normally expected to rapidly vanish with the order increasing, is a well-known fact that certain physical phenomena such as magnetism in solid $_3$He cannot be understood without inclusion of four-particle interaction terms [74,75]. It is worth here to mention that a detailed theory of optomechanics in superfluid $_4$He has been developed [76], but no expression for the nonlinear terms has been reported.

In general, the interaction of mechanical and optical modes is not strictly one-dimensional, implying that the overlap integral of normalized modes should also be taken into account. For instance, odd mechanical modes with even optical modes have zero interaction. In that sense, tuning the interaction to an odd mode and then shining an even optical mode, or vice versa, makes the optomechanical interaction identically zero by setting $\alpha \equiv \sqrt{2}g_0 = 0$. Then the lowest order surviving interaction would be the $\mathbb{H}_4$ term. This method has been used in [55-57] to highlight the quadratic interaction and make its measurement much easier. It has been shown that these quadratic terms may be exploited for direct observation of mechanical eigenmode jumps [55,56], as well as two-phonon cooling and squeezing [57], while the coupling strength $\beta$ could be increased by three orders of magnitude [56].

Moreover, the origin of mechanical parametric coupling which has recently been phenomenologically hypothesized [58] for the associated physical interactions cannot be understood without the presented analysis, although based on some earlier experimental evidence [77].

It must be added that the condition $\omega \gg \Omega$ may be violated in carefully designed superconducting microwave circuits and also the recently demonstrated molecular optomechanics [78], which signifies the importance of the $\mathfrak{P}^2\mathbb{Q}^2$ term in $\mathbb{H}_4$. It is furthermore worthwhile to point out that the regime $\omega = \Omega$ can be indeed be accessed and investigated, as it has been shown experimentally for superconducting circuit optomechanics [79]. The proposal of light propagation in a cylinder with rotating walls [80] also requires accessing regimes where $\omega$ and $\Omega$ fall within the same order of magnitude. Alternatively, in situations where $\omega \ll \Omega$, the scaling will be then given as

$$\theta = R\frac{\Omega^2 x_{zp}}{\omega^2 l}, \tag{49}$$

which shows a significant enhancement in this type of interactions.

*3.1. Conditions for Observation of Momentum-Field Quadratic Interactions*

In summary, two general criteria should be satisfied in a carefully designed experiment to allow investigation of momentum-field quadratic interactions:
- The optomechanical interaction $\mathbb{H}_3$ must vanish to allow easier study of quadratic interaction $\mathbb{H}_4$. This is quite possible by design as extensively has been discussed in the above and literature [55-62].

- The mechanical frequency $\Omega$ must be of the same order of magnitude or exceeding the electromagnetic frequency $\omega$. This is also possible and at least one experiment using superconducting optomechanics [79] has accessed this regime. Other possibilities are molecular optomechanics [78] as well as a rotating cylinder [80].

Evidently, such momentum-field quadratic interactions might be more difficult to observe under normal experimental conditions compared to the regular optomechanical setups. However, progressive developments in the precision and accuracy of optomechanics experiments, such as what happened for the case of Laser Interferometric Gravitational Observatory (LIGO) [81], could make it eventually possible to realize and probe such unexplored domains.

*3.2. Linearized Quantization*

3.2.1 Optical Field

The standard method to linearize the interaction Hamiltonian can be now used by making the substitutions $\hat{a} \to \bar{a} + \hat{a}$ where the new $\hat{a}$ operator from now on stands for the non-classical perturbations and $\langle \hat{a} \rangle$ is a measure of optical field amplitude. Then ignoring higher-order terms and retaining only the lowest-order interacting terms, we get

$$\mathbb{H}_3 = -\hbar g_3 (\hat{b}^\dagger + \hat{b})(e^{i\varphi}\hat{a}^\dagger + e^{-i\varphi}\hat{a}), \tag{50}$$

as well as

$$\mathbb{H}_4 = +\hbar g_4^+ (\hat{b}^\dagger + \hat{b})^2 (e^{i\varphi}\hat{a}^\dagger + e^{-i\varphi}\hat{a}), \omega \gg \Omega,$$

$$\mathbb{H}_4 = +\hbar g_4^- (\hat{b}^\dagger - \hat{b})^2 (\hat{a}^\dagger + \hat{a}), \omega \ll \Omega. \tag{51}$$

Here, $\varphi = \sphericalangle \bar{a}$ and the coupling frequencies are defined as

$$g_3 = \frac{\alpha}{\sqrt{2}} |\bar{a}| \equiv g_0 |\bar{a}| \equiv G, \qquad g_4^+ = \frac{\beta}{2} |\bar{a}| = \theta G, \qquad g_4^- = R \frac{\Omega^2}{\omega^2} g_4^+. \tag{52}$$

3.2.2 Mechanical Field

Following the same method to linearize the mechanical motions, with the replacement $\hat{b} \to \bar{b} + \hat{b}$ where the new $\hat{b}$ operator denotes the perturbations, gives rise to the expressions

$$\mathbb{H}_4 = +\hbar G_4^+ (\hat{b}^\dagger + \hat{b})(e^{i\varphi}\hat{a}^\dagger + e^{-i\varphi}\hat{a}), \qquad \omega \gg \Omega,$$

$$\mathbb{H}_4 = +\hbar G_4^- (\hat{b}^\dagger - \hat{b})(\hat{a}^\dagger + \hat{a}), \qquad \omega \ll \Omega, \tag{53}$$

where $\vartheta = \sphericalangle \bar{b}$ is set to zero without loss of generality, $G_4^+ = 2|\bar{b}|g_4^+ \cos\vartheta$, and $G_4^- = 2|\bar{b}|g_4^- \sin\vartheta$. In general, when $\omega \gg \Omega$ is violated, one would expect the momentum of mirror be coupled to the first quadrature of the radiation field. This type of interaction can be compared to the normal optomechanical interaction (50), in which the position is coupled to the first quadrature of the field.

3.2.3 Squeezing Hamiltonian

Without When the optical and mechanical frequencies do not differ by orders of magnitude so that neither $\omega \gg \Omega$ nor $\omega \ll \Omega$ hold, then the linearized Hamiltonian could be recast as

$$\mathbb{H}_4 = \hbar G_4^+ (\hat{b}^\dagger + \hat{b})(e^{i\varphi}\hat{a}^\dagger + e^{-i\varphi}\hat{a}) + \hbar G_4^- (\hat{b}^\dagger - \hat{b})(\hat{a}^\dagger + \hat{a}), \tag{54}$$

This can be written as

$$\mathbb{H}_4 = \hbar G(\hat{a}\mathbb{B}^\dagger + \hat{a}^\dagger \mathbb{B}) = \hbar G(\hat{b}\mathbb{A}^\dagger + \hat{b}^\dagger \mathbb{A}), \tag{55}$$

where

$$G_4 = \sqrt{G_4^+ G_4^-}, \qquad \widehat{\mathbb{A}} = \hat{a}^\dagger \sinh\rho + \hat{a}\cosh\rho, \qquad \widehat{\mathbb{B}} = \hat{b}^\dagger \cosh\rho + \hat{b}\sinh\rho,$$

$$\rho = \tanh^{-1}\left(\frac{G_4^+ - G_4^- e^{i\varphi}}{G_4^+ + G_4^- e^{i\varphi}}\right). \tag{56}$$

It is here to be noticed that $\widehat{\mathbb{B}}$ and $\widehat{\mathbb{A}}$ are in the standard form of Bogoliubov squeezing operator [54,55]. It may be noted that the equation

$$G_4^- = R\frac{\Omega^2}{\omega^2} G_4^+, \tag{57}$$

is actually a function of $\vartheta$ by definition of $G_4^+$ and $G_4^-$. Simplifying the above gives the expression for squeeze ratio as

$$\rho = \ln\left(\frac{\omega}{\sqrt{R\Omega}}\right) - i\frac{\varphi}{2}. \tag{58}$$

This shows that quadratic interactions give rise to squeezed mechanical or optical states unless $\omega = \sqrt{R}\Omega$ and of course $\varphi = 0$.

### 3.2.4 Special Case

As discussed above, the Hamiltonian $\mathbb{H}_3$ can be made identically zero [56,57,82,83] to access the quadratic interaction terms $\mathbb{H}_4$ directly. There is an interesting condition on the ratio of optical to mechanical frequencies, which could be sought here. Let

$$\omega = \sqrt{\eta R}\Omega, \tag{59}$$

in which $\eta$ is a constant to be determined later. This allows the $\mathbb{H}_4$ to be written as

$$\mathbb{H}_4 = 2\hbar\beta\left[-\frac{1}{\eta}\mathfrak{P}^2\mathbb{Q}^2 + \mathfrak{X}^2(\mathbb{P}^2 + \mathbb{Q}^2)\right]$$

$$= \hbar\beta\left[\frac{1}{2\eta}(\hat{b}^\dagger - \hat{b})^2(\hat{a}^\dagger + \hat{a})^2 + (\hat{b}^\dagger + \hat{b})^2(\hat{a}^\dagger\hat{a} + \hat{a}\hat{a}^\dagger)\right]. \tag{60}$$

Further expansion of results in (Appendix B)

$$\mathbb{H}_{4,int} = 2\hbar\beta|\bar{a}|\left[\left(1 + \frac{1}{2\eta}\right)\left(\hat{b}^{\dagger 2} + \hat{b}^2\right) + \left(1 - \frac{1}{2\eta}\right)\widehat{m}\right] \times \left(\hat{a}^\dagger + \hat{a}\right). \tag{61}$$

Then, for the choice of $\eta = \frac{1}{2}$, that is $\omega \cong 0.69\Omega$, one may reach the desired interaction quadratic Hamiltonian, linearized in the electromagnetic operators

$$\mathbb{H}_{4,int} = \hbar 2J\left(\hat{b}^{\dagger 2} + \hat{b}^2\right)\left(\hat{a}^\dagger + \hat{a}\right), \tag{62}$$

where the interaction rate is $J = \lambda = 2\beta|\bar{a}|$. When expanded in its four terms and after the replacement $\hat{c} = \frac{1}{2}\hat{b}^2$ (Appendix C), one may immediately recognize the Hamiltonian of the type

$$\mathbb{H}_{4,int} = \hbar J(\hat{c}\hat{a}^\dagger + \hat{c}^\dagger\hat{a}) + \hbar\lambda(\hat{c}^\dagger\hat{a}^\dagger + \hat{c}\hat{a}). \tag{63}$$

The first parenthesis represents the Hopping or Beam-Splitter term, while the second is normally referred to as the dissipation. Interestingly, the above could have been further linearized in mechanical operators to obtain

$$\mathbb{H}_{4,int} = i\hbar\mathcal{J}\left(e^{-i\vartheta}\hat{b}^\dagger + e^{i\vartheta}\hat{b}\right)\left(\hat{a}^\dagger - \hat{a}\right) = i\hbar\mathcal{J}\left(e^{i\vartheta}\hat{b}\hat{a}^\dagger - e^{-i\vartheta}\hat{b}^\dagger\hat{a}\right) + i\hbar\mathcal{J}\left(e^{-i\vartheta}\hat{b}^\dagger\hat{a}^\dagger - e^{i\vartheta}\hat{b}\hat{a}\right). \tag{64}$$

where $\mathcal{J} = 2J|\bar{b}|$. This latter form, may find application in non-reciprocal optomechanics [84].

### 4. Relativistic Considerations

As a final remark, the approximate nature of the Lagranian formulation by Law [5] has not been left unnoticed. It could be attributed first to the non-relativistic description of mirror's motion which ultimately ignores higher-order interactions, and then to the relativistic nature of radiation friction force and the associated Doppler shift [66]. As a result, in a subsequent paper by Cheung and Law [67], it has been made clear that the non-relativistic optomechanical Hamiltonian is correct only to the first-order in $\dot{q}$.

The nature of the relativistic corrections can be quite different, as follows:
- relativistic Doppler shift [66], which causes corrections in $\dot{q}/c$,

- relativistic correction in radiation pressure term [67], the lowest-order of which being proportional to $\dot{q}/c$,
- length contraction [68], due to the moving mirror boundary, resulting again in corrections as $\dot{q}/c$.

Not surprisingly, all these relativistic terms vanish in the limit of infinite light speed $c$. These altogether could be taken into account in a fully relativistic formulation of the Lagrangian and equations of motions for the mirror and optical field [85], which has been recently carried out in an extensive research by Castaños & Weder [68].

As shown in Appendix D, the total relativistic correction terms added to the Hamiltonian takes the form

$$\Delta \mathcal{H} = -\hbar(\hat{b}^\dagger - \hat{b})^2 \sum_k w_{kj} (\hat{a}_k^\dagger + \hat{a}_k)(\hat{a}_j^\dagger + \hat{a}_j). \tag{65}$$

For the single-mode cavity, $w = \chi_0 \pi \hbar d\Omega / 4mcl^2$, to be compared with $\beta = \hbar\omega/m\Omega l^2$ in (48). Hence, the relativistic correction to the quadratic Hamiltonian $\mathbb{H}_4$ is

$$-\frac{w}{\beta} = -\frac{\chi_0 \pi d\Omega^2}{4c\omega}. \tag{66}$$

Again, it is seen that when $\omega \gg \Omega$ is violated, the relativistic corrections might be quite significant. In any case, there is no relativistic correction to $\mathbb{H}_3$.

## 5. Conclusions & Future Work

The derivation of the optomechanical Hamiltonian has been carefully examined from the modal expansions, equations of motion, all the way to the Lagrangian, and ultimately the Hamiltonian and relativistic considerations. A set of correction terms to the nonlinear terms have been identified, which do not eliminate under any choice of canonical momenta. With the careful system design which allows $g_0 = 0$, these type of interactions are particularly interesting and now being actively pursued. It was shown that under these conditions one may expect coupling of mechanical momentum to the field position. Other sorts of interactions emerge under various conditions. In general, when the optical frequency is not much larger than the mechanical frequency, novel nonlinear interactions may appear.

A future work of the author [86] discusses a semi-analytical method based on the Langevin equations, which will enable easier study of quadratic interactions in quantum mechanical systems. This modified method of Langevin equations could in principle greatly simplify the study of quadratic and higher-order interactions, which would otherwise need either a full numerical solution using the master equation approach or full expansion unto the infinite set of orthogonal basis number kets.

**Appendix A. Equations of Motion**

In this section, we present the step-by-step details of the derivation of (9) from the previous equations, as it constitutes the most critical part of this article.

Starting from (8), one has first to rename the dummy index from $k$ to $j$, multiply both sides by $f_k$, and then take the inner product. This will yield the expression

$$-\sum \omega_j^2 Q_j (f_k|f_j) = \sum \ddot{Q}_j (f_k|f_j) - \frac{\dot{q}}{q} \sum \dot{Q}_j [(f_k|f_j) + 2(f_k|\kappa_j x|g_j)]$$

$$-\frac{\ddot{q}}{q} \sum Q_j \left[\frac{1}{2}(f_k|f_j) + (f_k|\kappa_j x|g_j)\right] + \frac{\dot{q}^2}{q^2} \sum Q_j \left[\frac{3}{4}(f_k|f_j) + 3(f_k|\kappa_j x|g_j) - (f_k|\kappa_j^2 x^2|g_j)\right]. \tag{A1}$$

Using (3), we trivially get

$$-\sum \omega_j^2 Q_j \delta_{kj} = \sum \ddot{Q}_j \delta_{kj} - \frac{\dot{q}}{q} \sum \dot{Q}_j [\delta_{kj} + 2\alpha_{kj}] - \frac{\ddot{q}}{q} \sum Q_j \left[\frac{1}{2}\delta_{kj} + \alpha_{kj}\right]$$

$$+ \frac{\dot{q}^2}{q^2} \sum Q_j \left[\frac{3}{4}\delta_{kj} + 3\alpha_{kj} - \beta_{kj}\right], \tag{A2}$$

which after rearrangement takes the form

$$\ddot{Q}_k = -\omega_k^2 Q_k + \frac{\dot{q}}{q}\dot{Q}_k + \frac{\ddot{q}}{2q}Q_k - \frac{3\dot{q}^2}{4q^2}Q_k + 2\frac{\dot{q}}{q}\sum \alpha_{kj}\dot{Q}_j + \frac{\ddot{q}}{q}\sum \alpha_{kj}Q_j$$
$$- \frac{\dot{q}^2}{q^2}\sum \left(3\alpha_{kj} - \beta_{kj}\right)Q_j. \tag{A3}$$

We may furthermore use $\alpha_{kj} = -\frac{1}{2}\delta_{kj} + g_{kj}$ and $\beta_{kj} = \left(\frac{1}{3}k^2\pi^2 - \frac{1}{2}\right)\delta_{kj} + h_{kj}$ from (4) to simplify and rewrite (A3) as

$$\ddot{Q}_k = -\omega_k^2 Q_k - \frac{3\dot{q}^2}{4q^2}Q_k + \frac{3\dot{q}^2}{2q^2}Q_k + \frac{\dot{q}^2}{q^2}\left(\frac{k^2\pi^2}{3} - \frac{1}{2}\right)Q_k + 2\frac{\dot{q}}{q}\sum g_{kj}\dot{Q}_j + \frac{\ddot{q}}{q}\sum g_{kj}Q_j$$
$$- \frac{\dot{q}^2}{q^2}\sum \left(3g_{kj} - h_{kj}\right)Q_j, \tag{A4}$$

which by plugging in the definition for $r_k$ from (10) takes the form

$$\ddot{Q}_k = -\omega_k^2 Q_k + r_k\frac{\dot{q}^2}{q^2}Q_k + 2\frac{\dot{q}}{q}\sum g_{kj}\dot{Q}_j + \frac{\ddot{q}}{q}\sum g_{kj}Q_j + \frac{\dot{q}^2}{q^2}\sum \left(h_{kj} - 3g_{kj}\right)Q_j. \tag{A5}$$

This is exactly the equation (9).

**Appendix B. Special Case**

Expansion of (60) results in

$$\mathbb{H}_4 = \hbar\beta \left[\frac{1}{2\eta}\left(\hat{b}^\dagger - \hat{b}\right)^2\left(\hat{a}^{\dagger 2} + \hat{a}^2\right) + \left(1 + \frac{1}{2\eta}\right)\left(\hat{m} + \frac{1}{2}\right)\left(\hat{n} + \frac{1}{2}\right) + \left(\hat{b}^{\dagger 2} + \hat{b}^2\right)\left(\hat{n} + \frac{1}{2}\right)\right]$$
$$= \hbar\beta \left[\frac{1}{2\eta}\left(\hat{b}^{\dagger 2} + \hat{b}^2\right)\left(\hat{a}^{\dagger 2} + \hat{a}^2\right) + \left(1 + \frac{1}{2\eta}\right)\left(\hat{m} + \frac{1}{2}\right)\left(\hat{n} + \frac{1}{2}\right) + \left(\hat{b}^{\dagger 2} + \hat{b}^2\right)\left(\hat{n} + \frac{1}{2}\right)\right.$$
$$\left. - \frac{1}{2\eta}\left(\hat{m} + \frac{1}{2}\right)\left(\hat{a}^{\dagger 2} + \hat{a}^2\right)\right], \tag{B1}$$

where $\hat{n} = \hat{a}^\dagger\hat{a}$ and $\hat{m} = \hat{b}^\dagger\hat{b}$ are respectively photon and photon number operators. Retaining only the interacting terms, gives the expression

$$\mathbb{H}_{4,int} = \hbar\beta \left[\frac{1}{2\eta}\left(\hat{b}^{\dagger 2} + \hat{b}^2\right)\left(\hat{a}^{\dagger 2} + \hat{a}^2\right) + \left(1 + \frac{1}{2\eta}\right)\hat{m}\hat{n} + \left(\hat{b}^{\dagger 2} + \hat{b}^2\right)\hat{n} - \frac{1}{\eta}\hat{m}\left(\hat{a}^{\dagger 2} + \hat{a}^2\right)\right]. \tag{B2}$$

In the limit $\omega \gg \Omega$ with $\eta \to \infty$, $\mathbb{H}_4$ as in (53) is recovered. With linearization of the electromagnetic field operators, and removing the non-interacting terms the following is found

$$\mathbb{H}_{4,int} = 2\hbar\beta \left[\frac{1}{2\eta}\left(\hat{b}^{\dagger 2} + \hat{b}^2\right)\left(\bar{a}^*\hat{a}^\dagger + \bar{a}\hat{a}\right) + \left(1 + \frac{1}{2\eta}\right)\hat{m}\left(\bar{a}\hat{a}^\dagger + \bar{a}^*\hat{a}\right)\right.$$
$$\left. + \left(\hat{b}^{\dagger 2} + \hat{b}^2\right)\left(\bar{a}\hat{a}^\dagger + \bar{a}^*\hat{a}\right) - \frac{1}{\eta}\hat{m}\left(\bar{a}^*\hat{a}^\dagger + \bar{a}\hat{a}\right)\right], \tag{B3}$$

By continuing the work on the linearized quadratic interaction one obtains the expression

$$\mathbb{H}_{4,int} = 2\hbar\beta|\bar{a}| \left[\frac{1}{2\eta}\left(\hat{b}^{\dagger 2} + \hat{b}^2\right)\left(e^{-i\varphi}\hat{a}^\dagger + e^{i\varphi}\hat{a}\right) + \left(1 + \frac{1}{2\eta}\right)\hat{m}\left(e^{i\varphi}\hat{a}^\dagger + e^{-i\varphi}\hat{a}\right)\right.$$
$$\left. + \left(\hat{b}^{\dagger 2} + \hat{b}^2\right)\left(e^{i\varphi}\hat{a}^\dagger + e^{-i\varphi}\hat{a}\right) - \frac{1}{\eta}\hat{m}\left(e^{-i\varphi}\hat{a}^\dagger + e^{i\varphi}\hat{a}\right)\right], \tag{B4}$$

which for $\varphi = 0$ simplifies to

$$\mathbb{H}_{4,int} = 2\hbar\beta|\bar{a}| \left[\left(1 + \frac{1}{2\eta}\right)\left(\hat{b}^{\dagger 2} + \hat{b}^2\right) + \left(1 - \frac{1}{2\eta}\right)\hat{m}\right] \times \left(\hat{a}^\dagger + \hat{a}\right). \tag{B5}$$

**Appendix C. Squared Annihilator**

The operator $\hat{c}$ has clearly a simple solution for its eigenkets, which is the same as coherent states such as $|z\rangle$ where $\hat{c}|z\rangle = \frac{1}{2}z^2|z\rangle$. Hence, the eigenvalue is simply the complex number $\frac{1}{2}z^2$. Meanwhile, one has

$$|z\rangle = e^{-\frac{1}{2}|z|^2} \sum_{m=0}^{\infty} \frac{z^m}{\sqrt{m!}} |m\rangle. \tag{C1}$$

It is furthermore easy to check that $[\hat{c}, \hat{c}^\dagger] = \widehat{m} + \frac{1}{2} = \hat{b}^\dagger\hat{b} + \frac{1}{2}$ [86]. When, the mean phonon number is $\langle\widehat{m}\rangle = \frac{1}{2}$, then $\langle[\hat{c}, \hat{c}^\dagger]\rangle = 1$, which is quite similar to the commutator $[\hat{b}, \hat{b}^\dagger] = 1$.

**Appendix D. Relativistic Correction**

The relativistic Lagrangian density for a light field with normal incidence to a fully reflective and non-compressible moving mirror, correct to the first-order in $\dot{q}$ and $\ddot{q}$, reads [68]

$$\frac{\partial \mathcal{L}}{\partial V} = \frac{1}{2}(\mathbf{E}\cdot\mathbf{D} - \mathbf{H}\cdot\mathbf{B}) + \frac{\Gamma^2\epsilon_0}{2}\chi|\mathbf{E}\cdot\hat{z} - c\mathrm{B}\mathbf{B}\cdot\hat{y}|^2, \tag{D1}$$

where $\mathbf{E} = -\hat{z}\frac{\partial}{\partial t}A$, $\mathbf{B} = \nabla\times(A\hat{z}) = -\hat{y}\frac{\partial}{\partial x}A$, $\mathbf{D} = \epsilon_0\mathbf{E}$, and $\mathbf{B} = \mu_0\mathbf{H}$. Furthermore,

$$\mathrm{B} = \frac{v}{c} = \frac{\dot{q}}{c} = \frac{p}{mc}, \qquad \Gamma = \frac{1}{\sqrt{1-\mathrm{B}^2}}, \tag{D2}$$

and $\chi$ is a dimensionless shape function independent of $v$, being zero outside mirror and relative susceptibility of the mirror's dielectric $\chi_0$ inside, and $\epsilon_0$ is the permittivity of vacuum. By expanding in the powers of B, this Lagrangian gives the first- and second-order corrections to the quadratic Hamiltonian density as

$$\frac{\partial}{\partial V}\Delta\mathcal{H} = \frac{\partial}{\partial V}\Delta\mathcal{H}^{(1)} + \frac{\partial}{\partial V}\Delta\mathcal{H}^{(2)} = -\frac{\partial}{\partial \mathrm{B}}\frac{\partial \mathcal{L}}{\partial V}\bigg|_{\mathrm{B}=0}\mathrm{B} - \frac{1}{2}\frac{\partial^2}{\partial \mathrm{B}^2}\frac{\partial \mathcal{L}}{\partial V}\bigg|_{\mathrm{B}=0}\mathrm{B}^2. \tag{D3}$$

Hence, one may obtain

$$\Delta\mathcal{H}^{(1)} = -\epsilon_0 S \int_0^q \frac{\partial}{\partial \mathrm{B}}\left[\frac{1}{2}\left(\frac{1}{\sqrt{1-\mathrm{B}^2}}\right)^2 \chi|A_t + c\mathrm{B}A_x|^2\right]\bigg|_{\mathrm{B}=0} \mathrm{B}\,dx, \tag{D4}$$

which further simplifies as

$$\Delta\mathcal{H}^{(1)} = -\epsilon_0 Sc \int_0^q \chi A_t A_x \mathrm{B}\,dx. \tag{D5}$$

It is appropriate to assume the approximation of conducting interface [87-89] for the mirror, such as the thickness is let to approach zero, while it susceptibility increases proportionally. In that limit, one may set

$$\chi(x,t) \approx \chi_0 d\delta[x - q(t)], \tag{D6}$$

where $d$ is the mirror's thickness. This is similar to the assumption of the locality of interaction by Gardiner & Zoller [90], too. Hence, one gets

$$\Delta\mathcal{H}^{(1)} \approx -\epsilon_0 V c\chi_0 A_t(q,t)A_x(q,t)\mathrm{B}, \tag{D7}$$

with $V = Sd$ is the cavity volume. Now, one has from (2), $A_x(q,t) = s\pi\sqrt{2/q^3}\sum kQ_k$, $A_t(q,t) = -\dot{q}A_x(q,t)$, and thus

$$\Delta\mathcal{H}^{(1)} \approx 2\pi^2 d\chi_0 \frac{\dot{q}^2}{q^3} \sum kj Q_k Q_j \tag{D8}$$

It is quite remarkable that (D8) is purely relativistic, and vanish in the limit of infinite $c$, as shown below. Here, the dependence on $t$ is hidden for convenience. This term translates after symmetrization into

$$\Delta\mathcal{H}^{(1)} = -2\hbar(\hat{b}^\dagger - \hat{b})^2 \sum_k w_{kj}(\hat{a}_k^\dagger + \hat{a}_k)(\hat{a}_j^\dagger + \hat{a}_j), \tag{D9}$$

where $w_{kj} = \sqrt{jk}\,\chi_0\pi\hbar d\Omega/4mcl^2$ are the coupling rates. Now, the quadratic correction $\Delta\mathcal{H}^{(2)}$ is given by

$$\Delta\mathcal{H}^{(2)} = -\frac{1}{2}\epsilon_0 S \int_0^q \frac{\partial^2}{\partial \mathrm{B}^2}\left\{\frac{1}{2}\left(\frac{1}{\sqrt{1-\mathrm{B}^2}}\right)^2 \chi|A_t + c\mathrm{B}A_x|^2\right\}dx \bigg|_{\mathrm{B}=0} \mathrm{B}^2. \tag{D10}$$

Simplifying and using the conducting interface approximation gives

$$\Delta \mathcal{H}^{(2)} = -\frac{1}{2} S \epsilon_0 \int_0^q \chi [A_t^2 + c^2 A_x^2] \mathrm{B}^2 dx \approx -\frac{1}{2} V \epsilon_0 \chi_0 [\dot{q}^2 + c^2] A_x^2(q,t) \mathrm{B}^2$$

$$\approx -\pi^2 d \chi_0 \frac{\dot{q}^2}{q^3} \sum kj Q_k Q_j. \quad \text{(D11)}$$

This one after insertion of operators gives $\Delta \mathcal{H}^{(2)} = -\frac{1}{2} \Delta \mathcal{H}^{(1)}$ and thus the total relativistic correction is found as

$$\Delta \mathcal{H}^{(1)} + \Delta \mathcal{H}^{(2)} = -\hbar (\hat{b}^\dagger - \hat{b})^2 \sum_k w_{kj} (\hat{a}_k^\dagger + \hat{a}_k)(\hat{a}_j^\dagger + \hat{a}_j). \quad \text{(D12)}$$